# A Time-Power Series Based Semi-Analytical Approach for Power System Simulation

Bin Wang, *Student Member, IEEE*, Nan Duan, *Student Member, IEEE*, and Kai Sun, *Senior Member, IEEE*

*Abstract*—Time domain simulation is the basis of dynamic security assessment for power systems. Traditionally, numerical integration methods are adopted by simulation software to solve nonlinear power system differential-algebraic equations about any given contingency under a specific operating condition. An alternative approach promising for online simulation is to offline derive a semi-analytical solution (SAS) and then online evaluate the SAS over consecutive time windows regarding the operating condition and contingency until obtaining the simulation result over a desired period. This paper proposes a general semi-analytical approach that derives and evaluates an SAS in the form of power series in time to approximate the solutions of power system differential equations. An error-rate upper bound of the SAS is also proposed to guarantee the reliable use of adaptive time windows for evaluation of the SAS. A dynamic bus method is proposed to extend the semi-analytical approach for solving general power system DAEs by efficiently linking the SASs for dynamic components through the numerical solution of the network algebraic equations. Case studies performed on the New England 39-bus system and the Polish 2383-bus system test the performance of the proposed semi-analytical approach and compare to existing methods. The results show that the SAS based approach has potentials for online simulations.

## I. INTRODUCTION

TIME domain simulation as the basis of dynamic security assessment (DSA) for power systems needs to solve power system differential-algebraic-equations (DAEs) for a list of credible contingencies occurring under a number of operating conditions. The increasing penetration of renewables and other intermittent energy resources are stressing transmission networks and adding uncertainties to grid operations. Power systems will experience more frequent disturbances and more diversified operating conditions than ever. In the next ten years, it is expected that DSA will transition from offline or day-ahead studies to the real time operation environment, i.e. the most credible contingencies are simulated under the real-time system condition estimated every 1-5 minutes by the state estimator. The power industry and the research community are looking forward to new methods enabling "faster-than-real-time simulation" [1], i.e. the clock time spent on a simulation run being shorter than the entire simulated time period.

To speed up the time domain simulation, parallel computing based on high performance computers are adopted [2]-[4]. At present, numerical integration methods such as Runge-Kutta, Trapezoidal and backward differentiation formula methods are still popular choices in most commercial simulation software. However, their sequential computation mechanisms do not make them fit directly in a parallel computing architecture. Thus, sophisticated schemes are specially designed to parallelize their computations. Papers [5][6] parallelize the numerical integration by means of the Parareal in time algorithm using a number of iterations on the decomposed small time intervals of the simulation period that are linked by a coarse solution over the whole period. Paper [7] applies the waveform relaxation approach to implement a parallelism through state variables. Other measures to speed up simulation are such as using an adaptive time window [8] and the network decomposition to enable parallel simulations on subsystems [9][10].

As an alternative approach, recent works [11]-[16] propose using a so-called semi-analytical solution (SAS) of power systems to perform fast simulations. As an approximate but explicit solution, an SAS is derived offline for only once for a range of probable system conditions and then evaluated online regarding the actual system condition and the contingency to be simulated. In the online stage, since the SAS is in the form of an explicit expression about time and system parameters, a simulation result can directly be obtained by replacing its symbolic variables by actual values. This SAS based approach is promising for online power system simulation because of the following advantages: (i) symbolizing selected system parameters to accommodate different system conditions, contingencies and uncertainties; (ii) requiring no iteration in SAS evaluation; (iii) fitting naturally in the parallel computing architecture because an SAS is in the form of a summation of terms, typically polynomials, and can be evaluated in parallel.

The SAS proposed in [11]-[13] is based on the Adomian Decomposition Method (ADM) [14]. A major drawback of the ADM-based SAS is that the offline derivation is very slow, especially for large-scale power systems. Also, the adaptive time window achieved in [13] is only based on the highest-order term of the derived SAS. Although a sharp change in the highest-order term may usually indicate the divergence of the SAS, a rigorous error-rate upper bound of the SAS has yet to be studied to guarantee a reliably adaptive time window. The

This work was supported in part by the ERC Program of the NSF and DOE under NSF grant EEC-1041877 and in part by NSF grant ECCS-1610025.

B. Wang, N. Duan and K. Sun are with the department of Electrical Engineering and Computer Science at the University of Tennessee, Knoxville, TN 37996 (e-mail:bwang13@vols.utk.edu, nduan@utk.edu, kaisun@utk.edu).

SAS in [15] is based on power series in time, which only considers the classical generator model and is only used for simulating the fault-on trajectories. Paper [16] uses Padé approximants to extend the time window of the SAS in [15] but does not provide an adaptive way to determine the time window. The above papers only consider power system models that can be represented by ordinary differential equations (DEs) by, e.g. reducing the passive power network and loads. However, more realistic power system models are usually represented by DAEs and need to consider dynamic loads.

Paper [17] applies a multi-stage ADM to simulate a power system modeled by DAEs. SASs are derived for dynamic elements modeled by DEs and are linked by numerical computations of algebraic equations (AEs) on the network. However, the length of time windows for computation is within 1ms and adaptive time windows are not applied. As reported in [17], solving the network AEs is the most time-consuming step and is performed for every time window.

This paper proposes a general semi-analytical approach that derives and evaluates an SAS in the form of power series in time to approximate the solution of power system DAEs. The paper also presents two schemes to implement the proposed semi-analytical approach: 1) it can be applied to a set of DEs that model the entire power system having AEs reduced such that a single SAS is derived for each state variable as a function of time, the initial state and system-wide parameters; 2) it can also be applied to a general power DAE model by finding SASs for individual dynamic elements and efficiently integrating evaluations of those SASs with the numerical solution of the network AEs using a proposed dynamic bus method and adaptive time windows.

The major contributions of this paper include: 1) a constructive proof of the existence of the time-power series based SAS for general nonlinear DEs, 2) fast offline derivation of SAS; 3) a proposal of the dynamic bus method extending the SAS based simulation from DEs to DAEs, 4) the proposal of error-rate upper bound for an SAS to determine the largest adaptive time window satisfying any given error-rate tolerance, 5) SAS based simulation using adaptive time windows, which solves the network AEs for fewer times leading to a less overall time cost of simulation, and 6) demonstration of the proposed semi-analytical approach on a realistic power system model with detailed generator models and dynamic load models considering motor loads.

As a result, with a longer time window in each simulation step, the total number of time steps for simulating a specific time length becomes less. Thus, the network AE is solved for fewer times and the overall time performance for simulating power system DAEs can be improved.

The rest of the paper is organized as follows. Section II proves the existence of the time-power series based SAS for general nonlinear DEs and applies the SAS for simulating power system DAEs with the proposed dynamic bus. Section III proposes an error-rate upper bound for determining a reliably adaptive time window for SAS-based time domain simulation. In Section IV, the accuracy of the proposed SAS and the performance of the proposed adaptive time windows are investigated on a simple linear system. Then, the time performance and accuracy of the SAS based simulation are investigated on the New England 39-bus system and the Polish 2383-bus system. Finally, conclusions are drawn in section V.

## II. TIME-POWER SERIES BASED SEMI-ANALYTICAL SOLUTION

### A. Time-power series based SAS for nonlinear DEs

Consider a general nonlinear dynamical system represented by a set of DEs

$$\begin{cases} \dot{\mathbf{x}}(t) = \mathbf{f}(\mathbf{x}(t)) & (1a) \\ \mathbf{x}(t_0) = \mathbf{x}_0 & (1b) \end{cases}$$

where $\mathbf{x}$ is the state vector of dimension $N \times 1$, $\mathbf{x}_0$ is the initial condition and $\mathbf{f}$ is a smooth vector field.

The existence and uniqueness of the exact solution $\mathbf{x}(t)$ to (1) can be guaranteed by Caratheodory's existence theorem [18]. Also, assume that $\mathbf{x}(t)$ is differentiable with respect to $t$ up to a desired order. Thus, $\mathbf{x}(t)$ can be expanded to the power series in (2) with coefficient vectors $\mathbf{a}$ to be determined. The time derivative of $\mathbf{x}(t)$ is given in (3). Note that $\mathbf{a}_0$ equals the initial state $\mathbf{x}_0$, and $\mathbf{a}_k$ ($k \geq 1$) depends on $\mathbf{a}_0$ and system parameters, denoted by vector $\mathbf{p}$.

$$\mathbf{x}(t) = \sum_{k=0}^{\infty} \mathbf{a}_k \cdot (t - t_0)^k \quad (2)$$

$$\dot{\mathbf{x}}(t) = \sum_{k=0}^{\infty} (k+1)\mathbf{a}_{k+1} \cdot (t - t_0)^k \quad (3)$$

We call the solution in (2) the time-power series based solution to the initial value problem (IVP) in (1). If those unknown coefficient vectors $\mathbf{a}$ can be solved analytically, then we call the solution in (2) the time-power series based SAS to the IVP in (1). Next we will show an algorithm which analytically solves for $\mathbf{a}$ in a recursive way.

**START**
1. Substitute (2) into right hand side of (1a) and find the Taylor expansion at $t = t_0$ as shown in (4).
2. Substitute (3) and (4) into (1a) and obtain (5) and (6).
3. Let $k = 0$.
4. Solve for $\mathbf{a}_{k+1}$ from (6), i.e. $\mathbf{a}_{k+1} = \mathbf{b}_k/(k+1)$.
   $k = k + 1$.
5. **go to 4**

$$\mathbf{f}(\mathbf{x}) = \mathbf{f}\left(\sum_{k=0}^{\infty} \mathbf{a}_k \cdot (t - t_0)^k\right) = \sum_{k=0}^{\infty} \mathbf{b}_k \cdot (t - t_0)^k \quad (4)$$

$$\sum_{k=0}^{\infty} (k+1)\mathbf{a}_{k+1} \cdot (t - t_0)^k = \sum_{k=0}^{\infty} \mathbf{b}_k \cdot (t - t_0)^k \quad (5)$$

$$(k+1)\mathbf{a}_{k+1} = \mathbf{b}_k \quad \text{for } k = 0, 1, 2, \cdots \quad (6)$$

The above algorithm is feasible only if $\mathbf{b}_k$ is known when solving for $\mathbf{a}_{k+1}$ in step 4, which is true according to the observation below.

Since $\mathbf{x}(t_0) = \mathbf{x}_0$, then the Taylor expansion of $\mathbf{f}(\mathbf{x}(t))$ at $t_0$ is equivalent to the Taylor expansion of $\mathbf{f}(\mathbf{x})$ at $\mathbf{x}_0$, i.e.

$$\mathbf{f}(\mathbf{x}(t)) = \mathbf{f}(\mathbf{x}) \Rightarrow \sum_{k=0}^{\infty} \mathbf{b}_k \cdot (t-t_0)^k = \mathbf{f}(\mathbf{x}_0) + \sum_{k=1}^{\infty} \mathbf{q}_k \cdot \Delta \mathbf{x}^k \quad (7)$$

$$\mathbf{q}_k = \left. \frac{\partial^k \mathbf{f}}{\partial \mathbf{x}^k} \right|_{\mathbf{x}=\mathbf{x}_0} \quad (8)$$

$$\Delta \mathbf{x} = \mathbf{x}(t) - \mathbf{x}_0 = \sum_{k=0}^{\infty} \mathbf{a}_k \cdot (t-t_0)^k - \mathbf{x}_0 = \sum_{k=1}^{\infty} \mathbf{a}_k \cdot (t-t_0)^k \quad (9)$$

where $\Delta \mathbf{x}^k$ consists of all $k^{\text{th}}$ order homogeneous polynomials in $\Delta \mathbf{x}$ and $\mathbf{q}_k$ is the corresponding coefficient matrix. Note that $\mathbf{q}_k$ does not depend on $t$ or any $\mathbf{a}_k$ with $k>0$.

Base on (9), how the coefficient of $(t-t_0)^m$ in $\Delta \mathbf{x}^n$ depends on the unknowns in $\mathbf{a}$ can be determined, as shown in Table I. For example, the element in the rectangular means that the coefficient of $(t-t_0)^2$ in $\Delta \mathbf{x}^1$ only depends on $\mathbf{a}_0$ and $\mathbf{a}_2$. Thus, considering (7), Table I shows that $\mathbf{b}_k$, i.e. the coefficient of $(t-t_0)^k$ on the left side of (7), only depends on $\mathbf{a}_0, \mathbf{a}_1, \ldots, \mathbf{a}_k$, for any $k=0, 1, 2,\ldots$. Thus, (10) holds, where $\mathbf{g}_k$ is a vector function depending only on $\mathbf{a}_0, \mathbf{a}_1, \ldots, \mathbf{a}_k$.

Finally, substitute (10) into (6) to give (11a), which is a recursive solution for coefficients $\mathbf{a}$. Note that by substitution, $\mathbf{a}_{k+1}$ only depends on $\mathbf{a}_0$ and $\mathbf{p}$, i.e. (11b).

TABLE I
COEFFICIENTS OF $(t-t_0)^m$ IN $\Delta \mathbf{x}^n$

| | $\Delta\mathbf{x}^0$ | $\Delta\mathbf{x}^1$ | $\Delta\mathbf{x}^2$ | $\Delta\mathbf{x}^3$ | $\Delta\mathbf{x}^4$ | ... | $\Delta\mathbf{x}^k$ | ... |
|---|---|---|---|---|---|---|---|---|
| $(t-t_0)^0$ | $\mathbf{a}_0$ | 0 | 0 | 0 | 0 | ... | 0 | ... |
| $(t-t_0)^1$ | 0 | $\mathbf{a}_0, \mathbf{a}_1$ | 0 | 0 | 0 | ... | 0 | ... |
| $(t-t_0)^2$ | 0 | $\mathbf{a}_0, \mathbf{a}_2$ | $\mathbf{a}_0, \mathbf{a}_1$ | 0 | 0 | ... | 0 | ... |
| $(t-t_0)^3$ | 0 | $\mathbf{a}_0, \mathbf{a}_3$ | $\mathbf{a}_0, \mathbf{a}_1, \mathbf{a}_2$ | $\mathbf{a}_0, \mathbf{a}_1$ | 0 | ... | 0 | ... |
| ⋮ | ⋮ | ⋮ | ⋮ | ⋮ | ⋮ | | ⋮ | ... |
| $(t-t_0)^k$ | 0 | $\mathbf{a}_0, \mathbf{a}_k$ | $\mathbf{a}_0\ldots\mathbf{a}_{k-1}$ | $\mathbf{a}_0\ldots\mathbf{a}_{k-2}$ | $\mathbf{a}_0\ldots\mathbf{a}_{k-3}$ | ... | $\mathbf{a}_0, \mathbf{a}_1$ | ... |
| ⋮ | ⋮ | ⋮ | ⋮ | ⋮ | ⋮ | | ⋮ | ... |

$$\mathbf{b}_k = \mathbf{g}_k(\mathbf{a}_0, \mathbf{a}_1, \cdots, \mathbf{a}_k) \quad \text{for } k=0,1,2,\cdots \quad (10)$$

$$\mathbf{a}_{k+1} = \frac{\mathbf{g}_k(\mathbf{a}_0, \mathbf{a}_1(\mathbf{p}), \cdots, \mathbf{a}_k(\mathbf{p}))}{k+1} \quad \text{for } k=0,1,2,\cdots \quad (11\text{a})$$

$$\mathbf{a}_{k+1} = \frac{\tilde{\mathbf{g}}_k(\mathbf{a}_0, \mathbf{p})}{k+1} \quad \text{for } k=0,1,2,\cdots \quad (11\text{b})$$

***Remark***: The above algorithm shows that the time-power series based SAS can always be derived in a recursive way for ordinary DEs. In addition, in practical implementation, only a finite number of polynomial terms in $t$ can be handled in (2), say $n$, which will lead to totally $N\times n$ equations in (6) used for solving $N\times n$ unknown coefficients $\mathbf{a}_1, \mathbf{a}_2, \ldots, \mathbf{a}_n$. Denote the corresponding time-power series based SAS as (12), which is an truncated approximation of the true solution $\mathbf{x}(t)$, where $n$ is the order of the SAS $\mathbf{x}_{sas}^{<n>}(t)$.

$$\mathbf{x}(t) \approx \mathbf{x}_{sas}^{<n>}(t) = \sum_{k=0}^{n} \mathbf{a}_k \cdot (t-t_0)^k \quad (12)$$

The proposed time-power series approximation is not the only form for an analytical approximant of the solution $\mathbf{x}(t)$. For instance, considerable work has been done that integrates harmonic components towards an approximate solution. Two well-established analytical approximation methods are the Krylov-Boguliubov method and the power series method documented in [19], which both approximate an exact solution as sinusoidal functions of time: the former yields an approximation in a decaying periodic form and the latter includes higher frequency components in its resulting approximant. Those two methods can be used for analytical studies on the dynamics of a simple nonlinear system.

However, for a complex nonlinear system like a large-scale power grid, analytical approximation of the exact solution for the entire system is not practical. Obtaining numerical solutions is usually the conventional approach. The approach proposed by this paper is a hybrid semi-analytical-semi-numerical approach, whose approximant solution x(*t*) in the form of time-power series has advantages, especially in the presence of AEs. For DAEs, approximating both state variables and algebraic variables as time-power series provides a more natural interfacing scheme.

It should be noted that the approach proposed in this paper can also store the SAS as [13] does by using (11b) if offline deviation speed of the SAS is not a concern. For a large power system with many generators, the power series expression (12) could be very long even if *n* is not big. When parallel computing is allowed, the online evaluation speed of an SAS largely depends on how the SAS is stored in the memory. The ADM in [13] proposes to store the SAS with all coefficients of time directly expressed by system parameters $\mathbf{p}$ and initial state $\mathbf{x}_0$, the same as in (11b), which may require huge memory for large systems. This paper proposes to utilize the above recursive derivation, i.e. (11a), of the SAS $\mathbf{x}_{sas}^{<n>}$ such that *n* sequential steps are required to evaluate coefficients $\mathbf{a}_1, \ldots, \mathbf{a}_{n-1}$ and $\mathbf{a}_n$, respectively, where each step can be implemented using the parallel computing. Such sequential evaluation of an SAS will save tremendous memory for storing the SAS and significantly speed up its offline derivation but slightly sacrifice the upper limit of online evaluation speed (see the case study in section IV-B).

This subsection derived time-power series based SASs for a power system modeled by DEs. The SAS of each state variable is a function of time, the initial states and system parameters. For small power systems with, e.g., tens of generators, the SAS can be directly evaluated leveraged by parallel computing to speed up the simulation for online performance requirements [13][15][16]. However, for large power systems, tremendous storage and computing resources are required for deriving and evaluating an SAS for online applications. The next subsection will extend the semi-analytical approach to a hybrid approach for general large power systems modeled by DAEs. The hybrid approach solves DEs of dynamic elements using SASs and solves the network AEs by numerical iterations.

*B. Time-power series based SAS for power system DAEs*

A realistic power system model is usually represented by a large set of DAEs, including DEs for, e.g., generators, associate controllers and dynamic loads, etc., and AEs for the network and static loads. Directly finding the symbolized time power series based SAS for each state variable of such a system is difficult (though not theoretically impossible) mainly because of calculating inverses of high-dimensional symbolic matrices.

This section proposes a *dynamic bus* method to extend the semi-analytical approach for simulating large power systems represented by DAEs. For an SAS, a *dynamic bus* is used to represent a selected bus whose bus voltage is expressed by an explicit function of time and selected parameters to be continuously updated through simulation. The function should be selected to accurately represent the dynamics of the bus voltage over a period of time, covering each evaluation time window of the SAS. In this paper, the form of time-power series is selected for each bus voltage. The basic idea of the dynamic load method is introduced in this subsection while detailed steps of the SAS-based simulation involving such dynamic buses will be presented in next subsection.

For a general *g*-generator, *b*-bus, *l*-load, *m*-motor power system, denote the sets of all generator, load and motor buses respectively as $\mathcal{B}$, $\mathcal{G}$, $\mathcal{L}$ and $\mathcal{M}$. Obviously, there are $g \leq b$, $l \leq b$, $m \leq l$. Without loss of generality, consider a 6th order generator model, a 1st order exciter, a 1st order governor and the load model with ZIP load plus a 3rd order motor load, which are respectively shown in (13)-(17) [20][21]. The network equation is shown in (23). The initialization of these models can be found in [21]. The dynamic load method can also be applied to more complex power system models.

$$\begin{cases} \dot{\delta}_i = \Delta\omega_i \cdot \omega_s \\ \Delta\dot{\omega}_i = (P_{mi} - P_{ei} - D_i\Delta\omega_i)/2H_i \\ \dot{e}'_{qi} = \left(-\dfrac{X_{di} - X''_{di}}{X'_{di} - X''_{di}} \cdot e'_{qi} + \dfrac{X_{di} - X'_{di}}{X'_{di} - X''_{di}} \cdot e''_{qi} + e_{fdi}\right)\Big/T'_{d0i} \\ \dot{e}'_{di} = \left(-\dfrac{X_{qi} - X''_{qi}}{X'_{qi} - X''_{qi}} \cdot e'_{di} + \dfrac{X_{qi} - X'_{qi}}{X'_{qi} - X''_{qi}} \cdot e''_{di}\right)\Big/T'_{q0i} \\ \dot{e}''_{qi} = \left(e'_{qi} - e''_{qi} - (X'_{di} - X''_{di}) \cdot I_{di}\right)\big/T''_{d0i} \\ \dot{e}''_{di} = \left(e'_{di} - e''_{di} + (X'_{qi} - X''_{qi}) \cdot I_{qi}\right)\big/T''_{q0i} \end{cases}, \quad i \in \mathcal{G} \quad (13a)$$

$$\dot{e}_{fdi} = \left(K_i(V_{refi} - |V_i|) - e_{fdi}\right)/T_{ei}, \quad i \in \mathcal{G} \quad (13b)$$

$$\dot{P}_{mi} = \left(P_{refi} - P_{mi} - \Delta\omega_i/R_i\right)/T_{gi}, \quad i \in \mathcal{G} \quad (13c)$$

$$\begin{aligned} V_{di} + jV_{qi} &= e^{-j(\delta_i - \pi/2)}V_i \\ I_{di} &= \left((e''_{di} - V_{di})R_{ai} + (e''_{qi} - V_{qi})X''_{qi}\right)/\left(R_{ai}^2 + X''_{di}X''_{qi}\right) \\ I_{qi} &= \left(-(e''_{qi} - V_{qi})X''_{di} + (e''_{di} - V_{di})R_{ai}\right)/\left(R_{ai}^2 + X''_{di}X''_{qi}\right) \\ I_{gi} &= e^{j(\delta_i - \pi/2)}\left(I_{di} + jI_{qi}\right) \\ P_{ei} &= V_{di}I_{di} + V_{qi}I_{qi} \end{aligned} \quad (14)$$

where $\delta_i$, $\Delta\omega_i$, $e'_{qi}$, $e'_{di}$, $e''_{qi}$, $e''_{di}$, $e_{fdi}$ and $P_{mi}$ are state variables, representing rotor angle, rotor speed deviations, q- and d-axes transient and sub-transient field voltages, field voltage and the mechanical power of generator *i*; $V_{di}$, $V_{qi}$, $I_{di}$, $I_{qi}$ and $P_{ei}$ are d- and q-axes voltages and currents and the electrical power; $V_i$ is the terminal bus voltage, $I_{gi}$ is the current injected to the terminal bus from generator *i*; all others are parameters.

$$\begin{cases} \dot{s}_i = (T_{loadi} - T_{motori})/2H_{mi} \\ \dot{v}'_{di} = -\omega_s r_{ri}\left((X_{si} - X'_{si})I_{qmi} + v'_{di}\right)/X_{ri} + \omega_s s_i v'_{qi}, \quad i \in \mathcal{M} \\ \dot{v}'_{qi} = \omega_s r_{ri}\left((X_{si} - X'_{si})I_{dmi} - v'_{qi}\right)/X_{ri} - \omega_s s_i v'_{di} \end{cases} \quad (15)$$

$$\begin{aligned} y_i &= s_i X_{ri}/r_{ri} \\ z_{rei} &= r_{si} + y_i(X_{si} - X'_{si})/(1 + y_i^2) \\ z_{imi} &= X'_{si} + (X_{si} - X'_{si})/(1 + y_i^2) \\ I_{dmi} + jI_{qmi} &= V_i/(z_{re} + jz_{im}) \\ T_{motori} &= v'_{di}I_{dmi} + v'_{qi}I_{qmi} \\ T_{loadi} &= f_1 s_i^{i_1} + f_2(1 - s_i)^{i_2} \end{aligned} \quad (16)$$

where $s_i$, $v'_{qi}$ and $v'_{di}$ are state variables, representing motor slip, q- and d-axes transient voltages, respectively, of motor *i*.

$$I_{cci} + I_{cpi} = \dfrac{p_{i2}P_{i0}\dfrac{|V_i|}{|V_{i0}|} - jq_{i2}Q_{i0}\dfrac{|V_i|}{|V_{i0}|} + p_{i3}P_0 - jq_{i3}Q_0}{V_i^*} \quad (17)$$

$i \in \mathcal{L}$

where $I_{cci}$ and $I_{cpi}$ represent the currents from constant-current load and constant-power load at bus *i*, while the constant-impedance load is included into the network admittance matrix; $V_{i0}$, $P_{i0}$ and $Q_{i0}$ are the initial bus voltage, active and reactive loads at bus *i* at the initialization of the simulation; $p_{i2}$, $q_{i2}$, $p_{i3}$ and $q_{i3}$ are the percentages of constant-current and constant-power components of the active and reactive loads at bus *i*. Note that for load buses with and without a motor, we respectively have $p_{i1}+p_{i2}+p_{i3}+p_{im}=1$ and $p_{i1}+p_{i2}+p_{i3}=1$ for active load, where $p_{i1}$ and $p_{im}$ represent the percentages of constant-impedance load and motor load. We have similar equations in *q* for reactive power load.

$$\mathbf{I}_{\mathcal{B}} = \mathbf{Y}_{\mathcal{B}} \mathbf{V}_{\mathcal{B}} \quad (18)$$

where $\mathbf{I}_{\mathcal{B}}$ and $\mathbf{V}_{\mathcal{B}}$ are bus injection current and bus voltage vectors, $\mathbf{Y}_{\mathcal{B}}$ is the admittance matrix including the constant-impedance load. In this work, since each generator is interfaced with the network as a current source, therefore $\mathbf{Y}_{\mathcal{B}}$ in (18) also contains the admittances of the shunt branch of the equivalent current source of each generator.

The generator equations (13)-(14), motor equations (15)-(16) and the ZIP load equation (17) are linked through the network equation in (18). Seen from each dynamic element, the impacts from all other elements of the system is completely reflected by the dynamics of its terminal bus voltage, i.e. $V_i(t)$. If the terminal bus is considered a dynamic bus whose voltage can be expressed as a function $V_i(t)$ at least for a period of time, all dynamic elements will be decoupled individually from the rest of the system and hence they can all be modeled by DEs, whose time-power series based SAS can always be solved in a recursive way as shown in II-A. This is the main idea of the proposed dynamic bus method to extend the semi-analytical approach. Note that paper [17] also proposes to decouple dynamic elements from the network but the terminal bus is assumed to have a constant voltage in each

SAS. Thus, the method in [17] can be regarded as a special case of the proposed dynamic bus method.

In this paper, $V_i(t)$ is assumed to be in the form of a time-power series. The polar form is adopted since which is reported to outperform the rectangular form in terms of the extrapolation accuracy [22].

$$V_i(t) = V_{mi}(t)e^{jV_{ai}(t)} \approx \left(\sum_{k=0}^{n_v} V_{mik}t^k\right) e^{j\left(\sum_{k=0}^{n_v} V_{aik}t^k\right)} \quad (19)$$

To sum up, with the dynamic bus method, the SAS of the element $j$ on bus $i$ can be written as (20), where element $j$ can be a generator or a motor.

$$\mathbf{x}_{ij}(t, \mathbf{x}_{ijt0}, V_i, \mathbf{p}_{ij}) = \mathbf{x}_{ijt0} + \sum_{k=1}^{n} a_k (t-t_0)^k \quad (20)$$

where $\mathbf{x}_{ijt0}$ is the initial condition, $\mathbf{p}_{ij}$ represents the parameter set of the element $j$ on bus $i$, $a_k$ is the coefficient, which is a function of $\mathbf{x}_{ijt0}$, $\mathbf{p}_{ij}$, $V_{mi}$ and $V_{ai}$.

### C. SAS based simulation of power system DAEs

By means of the dynamic bus method, the semi-analytical approach can simulate a large power system modeled by DAEs following the flow chart shown in Fig. 1.

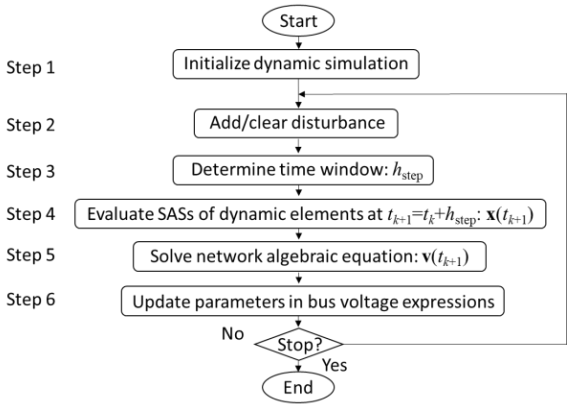

Fig. 1. Flow chart of SAS based simulation using the dynamic bus method.

There are following detailed steps:

**Step 1:** All dynamic elements are initialized at the stable equilibrium point of the pre-disturbance system.

**Step 2:** Disturbances can be added or cleared throughout the simulation, such as the disconnection or closure of a line, generation trip and load shedding. Note that any system conditions represented by the same set of DAEs with only differences in parameter values actually share the same symbolic SAS expression. For example, any 3-phase fault or any change in network topology will only change the admittance matrix, so exactly the same SAS can be used to simulate those disturbances by choosing proper parameters.

**Step 3:** The length of the time window for SAS evaluation can be either fixed, e.g. 1ms, or adaptive through simulation. An adaptive time window will be proposed in the next section.

**Step 4:** By the dynamic bus method, the SAS evaluations for all dynamic elements connected to either the same or different dynamic buses become independent of each other and hence can be parallelized on parallel computers.

**Step 5:** The network equation is solved through a few iterations to match the bus current injection in power flow model and the net current from the generators and loads on each individual bus. In each iteration, $\mathbf{V}_\mathcal{B}$ has to be solved from (18). In this work, the column approximate minimum degree permutation and the LU factorization are adopted to speed up the solution by making use of the sparsity of $\mathbf{Y}_\mathcal{B}$.

**Step 6:** The coefficients of all bus voltages in (19) are updated by solving a number of linear least square error problems using several bus voltage samples from previous time windows. Note that at the beginning of the simulation or after any disturbance, previous bus voltage samples may either be not available or not capable for the prediction to be used in the SAS. In this case, other numerical integration methods can always be adopted to start or restart the SAS based simulation. However, to achieve a self-starting SAS based simulation without resorting to other methods, a number of shorter sub-time windows are used to create enough samples for the estimation of coefficients in (19), where the SAS evaluations on these very short sub-time windows assume the bus voltages to be constant, i.e. $V_{mik} = V_{aik} = 0$ for $k \geq 1$, and the network AEs are solved in each sub-time window.

The proposed SAS based simulation using the dynamic bus method considers the dynamics of terminal bus voltages in each time window. As a comparison, paper [17] also adopts a similar partitioned solution approach, which assumes bus voltages and also a few other selected state variables to be constant in each time window. In fact, most of these variables are always continuously changing over time when the system is disturbed. Assuming them to be constant will inevitably reduce the radius of convergence (ROC) of the SASs of dynamic elements and lead to a limitation that a small time window, e.g. 1ms, has to be used for each SAS evaluation (i.e. simulation). In this paper, the dynamic bus method extends the ROC of each SAS so that a longer time window can be used.

*Remark:* the proposed SAS based simulation of power system DAEs is also able to further address the three factors below, which make it a promising candidate for online applications. Future investigations on these benefits is out of the scope of this paper.

**More complex models.** The proposed dynamic bus method enables the derivation of a time-power series based SAS for any dynamic element using a smooth function, including wind and solar generation and the saturation effect with a generator. This is because, with the terminal bus voltage expressed as a time function, dynamics of any element connected to that bus will be determined and its underlying equations become a set of DEs. Then, its time-power series based SAS can always be derived as shown in II-A. In addition, simulation considering stochastic effects of load and renewable generation can also be achieved by incorporating stochastic processes into the SAS thanks to its symbolic form [23].

**Changes in network topology.** Any changes in the network topology, e.g. the disconnection or closure of a transmission line, is reflected by using a different set of values in the admittance matrix $\mathbf{Y}_\mathcal{B}$ in (18). This does not have any impact on the derived SAS in a symbolic form, since $\mathbf{Y}_\mathcal{B}$ is symbolized and ready to take any values without changing the form of the SAS [13].

**Limits in controllers**. The limits in generator exciters or other controllers can be considered in the SAS based simulation by symbolizing necessary intermediate variables whose limits need to be addressed in the SAS expression and adding a limit checking procedure after the evaluation over each time window. Once any of such variables meet its upper or lower limit, its value will be fixed at the limit value in following time windows unless the variable is found to depart from the limit by the limit checking procedure.

## III. PROPOSED ADAPTIVE TIME WINDOW FOR SAS-BASED TIME DOMAIN SIMULATION

An SAS can be accurate, i.e. satisfying a given tolerance, within a short period of time. As an analytical, approximate solution, an SAS can also provide important information to determine the largest time window for simulations [13], which is related to the ROC problem. As a first step of finding the best approach having the largest ROC, this section presents a determination of the largest adaptive time window based on an proposed error-rate upper bound, for any given SAS to satisfy any given error tolerance. It is worth mentioning that the proposed method for determination of an adaptive time window works for other SASs, including those in [13][16][17].

### A. Determination of an error-rate upper bound

The term "error rate" represents the error of an approximate solution accumulated per unit time. Suppose that an SAS has been obtained, e.g. (12). Denote the time derivative of the SAS to be (21) and define function $r^{<n>}(h)$ as in (22), where $h \geq t_0$. Then, we have Theorem 1.

$$\dot{\mathbf{x}}_{sas}^{<n>}(t) = \sum_{k=1}^{n} k \mathbf{a}_k \cdot (t-t_0)^{k-1} \quad (21)$$

$$r^{<n>}(h) \triangleq \sup_{t \in [t_0, h]} \left\| \dot{\mathbf{x}}(t) - \dot{\mathbf{x}}_{sas}^{<n>}(t) \right\| = \sup_{t \in [t_0, h]} \left\| \mathbf{f}(\mathbf{x}(t)) - \dot{\mathbf{x}}_{sas}^{<n>}(t) \right\| \quad (22)$$

where $\|.\|$ represents a certain norm, e.g. infinity norm.

**Theorem 1.** Function $r^{<n>}(h)$ is an error-rate upper bound of the SAS $\mathbf{x}_{sas}^{<n>}$.

*Proof*. The error of the SAS $\mathbf{x}_{sas}^{<n>}$ accumulated from $t_0$ to the time instance $h$, is

$$\left\| \int_{t_0}^{h} \left( \dot{\mathbf{x}}(\tau) - \dot{\mathbf{x}}_{sas}^{<n>}(\tau) \right) d\tau \right\| = \left\| \int_{t_0}^{h} \left( \mathbf{f}(\mathbf{x}(\tau)) - \dot{\mathbf{x}}_{sas}^{<n>}(\tau) \right) d\tau \right\|$$
$$\leq \int_{t_0}^{h} \left\| \mathbf{f}(\mathbf{x}(\tau)) - \dot{\mathbf{x}}_{sas}^{<n>}(\tau) \right\| d\tau \quad (23)$$
$$\leq \int_{t_0}^{h} r^{<n>}(h) d\tau = r^{<n>}(h) \cdot (h-t_0)$$

Thus, here is (24) indicating $r^{<n>}(h)$ to be an error-rate upper bound. ∎

$$\frac{\left\| \int_{t_0}^{h} \left( \dot{\mathbf{x}}(\tau) - \dot{\mathbf{x}}_{sas}^{<n>}(\tau) \right) d\tau \right\|}{h-t_0} \leq r^{<n>}(h) \quad (24)$$

### B. Adaptive time window for SAS based simulation

This subsection will utilize the proposed error-rate upper bound to design an adaptive time window for the SAS based simulation.

The SAS is accurate at the initial condition and its error is assumed to monotonically increase over time within a small neighborhood of the initial time, as illustrated in Fig. 2, one can always find a time instance, say $h_{max}$, such that the error-rate upper bound is equal to a pre-specified error-rate tolerance, say $\varepsilon$. Then, the adaptive time window $h_{step}$ for the current simulation step is determined by $h_{step} = h_{max} - t_0$. For simplicity, $r^{<n>}(h)$ is denoted as $r(h)$ for the rest of this paper without ambiguity, where the order of the associate SAS will be clearly mentioned.

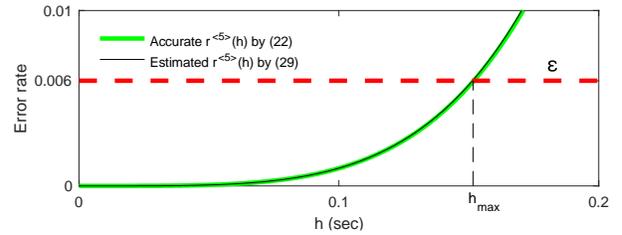

Fig. 2. Typical curve of the proposed error-rate upper bound. This plot is from the tested case in section IV-A with parameters $t_0 = 0$, $n = 5$ and $\varepsilon = 0.006$. Since the closed-form solution exists in this case, so accurate $r^{<5>}(h)$ can be calculated and compared with the estimated one, which almost overlap with each other. The adaptive time window determined in this step is about 0.152s.

In the implementation, the time cost for accurately finding $h_{max}$ is usually expensive due to the fact that the evaluation of $r(h)$ has to be done for many times. This may undermine the benefit from using an adaptive time window for increasing the simulation speed. Thus, an approximate but computationally efficient way is proposed for any given error-rate tolerance $\varepsilon$.

**Step 1**: Let the time window determined in previous step be $h_{pre}$. For the initial simulation step, $h_{pre}$ can take a small value, e.g. $h_{pre} = 1$ ms is used in this paper. In the current time window, calculate the SAS at $t_0+\alpha h_{pre}$ and the error-rate upper bound $r(t_0+\alpha h_{pre})$, where $\alpha$ is a safety factor adopted to maintain the conservativeness. $\alpha = 0.95$ is used in this paper.

**Step 2**: Approximate $r(h)$ by $\hat{r}(h)$ in (25). Use the point from step 1 to solve for $\mu$ as shown in (26).

**Step 3**: Approximate $h_{max}$ by $\hat{h}_{max}$ in (27).

**Step 4**: Calculate the adaptive time window $h_{step}$ by (28) to be used by the next time window.

$$\hat{r}(h) = \mu \left( e^{h-t_0} - 1 \right) \quad (25)$$

$$\mu = \frac{r(t_0 + \alpha h_{pre})}{e^{\alpha h_{pre}} - 1} \quad (26)$$

$$\hat{h}_{max} = t_0 + \ln\left( \varepsilon \frac{e^{\alpha h_{pre}} - 1}{r(t_0 + \alpha h_{pre})} + 1 \right) \quad (27)$$

$$h_{step} = \hat{h}_{max} - t_0 = \ln\left( \varepsilon \frac{e^{\alpha h_{pre}} - 1}{r(t_0 + \alpha h_{pre})} + 1 \right) \quad (28)$$

## C. Other Remarks

There are several important remarks for the proposed error-rate upper bound:

**Practical error-rate upper bound.** The proposed error-rate upper bound $r^{<n>}(h)$ in (22) is applicable to any SAS with the help of the assumed analytical solution. However, the exact solution does not exist for most systems requiring simulations. Thus, $r^{<n>}(h)$ does not have an explicit formulation in $h$. To make the proposed upper bound applicable in practice, the exact solution $\mathbf{x}(t)$ used in the definition (22) is replaced by the SAS $\mathbf{x}_{sas}^{<n>}(t)$, leading to an approximate error-rate upper bound.

$$r(h) \approx \sup_{t\in[t_0,t_0+h]} \left\| \mathbf{f}\left(\mathbf{x}_{sas}^{<n>}(t)\right) - \dot{\mathbf{x}}_{sas}^{<n>}(t) \right\| \quad (29)$$

**Conservativeness.** The proposed error-rate upper bound describes the upper bound of the SAS's error accumulated over a time interval. For a specific time window, since the actual error rate is not always equal to such an upper bound throughout the entire time interval, the actual accumulated error is often smaller than that predicted by the upper bound. Thus, using the proposed upper bound may usually result in a conservative, i.e. shorter than the actual, adaptive time window.

**Flexibility.** The proposed error-rate upper bound can be applied either to each single state variable to make sure the error of a specific quantity, e.g. rotor angle, within a pre-specified error rate tolerance, e.g. 0.01 degree per second, or to the whole state vector to guarantee some overall accuracy.

**Self-sufficiency.** The proposed approximate error-rate upper bound in (29) is self-sufficient, which means that for any derived SAS, its error rate can be bounded without resorting to any other SAS or numerical integration method.

**Error tolerance, simulation accuracy, SAS order and adaptive time window.** In the proposed SAS-based simulation using adaptive time window, the accuracy of the simulation depends almost entirely on the error tolerance. With the same error tolerance, SASs with different orders may use different time windows in their simulation according to the proposed adaptive time window scheme. Especially, using an SAS with a higher order would generally lead to a longer adaptive time window.

## IV. CASE STUDIES

This section first use a simple linear dynamical system having the closed-form solution to compare the proposed SAS based simulation with existing numerical integration methods, e.g. the 4th order Runge-Kutta (RK4) and backward differentiation formula (BDF) methods, in terms of the accuracy and variable time windows. Then, the proposed SAS is first applied to the New England 39-bus system [13] to demonstrate its performance on a power system modeled by DEs and then tested on the Polish 2383-bus power system [24] to demonstrate its accuracy and speedup when simulating a large power system modelled by DAEs.

## A. SAS vs. RK4 and BDF

In order to have the true solution as the reference to gain some insights on how the proposed SAS based simulation performs with adaptive time windows, we first test a linear dynamical system in (30) with a closed-form solution $x_{true}(t)$ in (31), and compare the SAS-based simulation to simulation results from the RK4 and BDF. The Matlab functions "ode45" and "ode15s" with specified options are used to simulate with RK4 and BDF, respectively. The time-power series based $N$-th order SAS is shown in (32), where initial state determines $a_0$ and $a_1$ by $a_0=x(0)$, $a_1=\dot{x}(0)$ and $a_k$ is determined recursively by (33) for $k=2, 3, \ldots, N$. Let the numerical solutions by the RK4 and BDF be $x_{rk4}(t)$ and $x_{bdf}(t)$, respectively. The error at time $t$ is defined in (34), where $x(t)$ could be any solution from SAS, RK4 or BDF. For all tests in this subsection, the parameters and initial conditions are taken as $\omega = \pi$, $\sigma = -0.1$, $x(0) = 0$, $\dot{x}(0) = \pi$.

$$\ddot{x} - 2\sigma\dot{x} + (\omega^2 + \sigma^2)x = 0 \quad (30)$$

$$x_{true}(t) = e^{\sigma t} \sin \omega t \quad (31)$$

$$x_{sas}^{<N>}(t) = \sum_{k=0}^{N} a_k t^k \quad (32)$$

$$a_k = \frac{2\sigma(k-1)a_{k-1} - (\omega^2+\sigma^2)a_{k-2}}{k(k-1)} \quad (33)$$

$$e(t) = x(t) - x_{true}(t) \quad (34)$$

The first test investigates the relationship between the accuracy and the order of the SAS when a fixed time window is used for evaluation, e.g. 0.01s is used here. Fig. 3 shows the 2nd and 3rd order SASs and their comparisons to the closed form solution, where the largest absolute errors are about 0.1 and $0.96 \times 10^{-3}$, respectively. Note that under the same condition, the largest absolute error of the RK4 is about $1.3 \times 10^{-3}$, comparable to that of the 3rd order SAS. Errors with higher-order SASs become even smaller as shown in Table II.

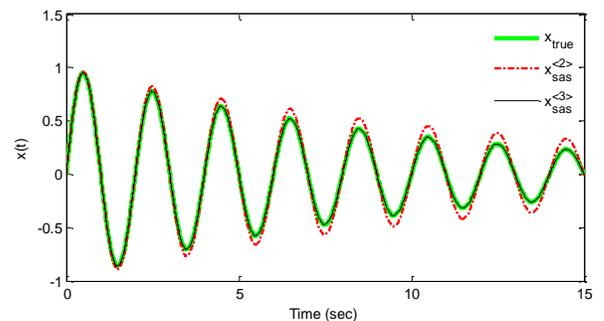

Fig. 3. Closed-form solution compared with 2nd and 3rd order SASs

TABLE II
LARGEST ABSOLUTE ERRORS OF SASs IN TIME DOMAIN

| $N$ | 4 | 5 | 6 | 7 | 8 |
|---|---|---|---|---|---|
| $\max\{|e_{sas}^{<N>}(t)|\}$ | 7.4e-6 | 4.7e-8 | 2.4e-10 | 1.3e-12 | 2.0e-13 |

The second test investigates the performance of the adaptive time window used in the proposed SAS based simulation. To make fair comparison to the RK4, the error tolerances with the BDF and SAS are carefully adjusted to make their errors be comparable to the default Matlab solver "ode45". For the BDF, it is found that setting option "AbsTol" to be $4\times10^{-5}$ with solver ode15s can achieve this. For simulations using the 3rd order to 8th order SASs, the tolerances of error rate, i.e. $\varepsilon$, need to take 0.0025, 0.004, 0.006, 0.007, 0.009 and 0.015, respectively. For example, the absolute errors calculated by (39) for three methods are shown in Fig. 4, which are all around the order of 0.001. Then, the fixed time window used by RK4 and the adaptive time windows by BDF and SASs are shown in Fig. 5, which shows: 1) the BDF and a 4th order SAS have comparable lengths of adaptive time windows; 2) SASs with orders higher than 4 can adopt longer adaptive time windows than BDF.

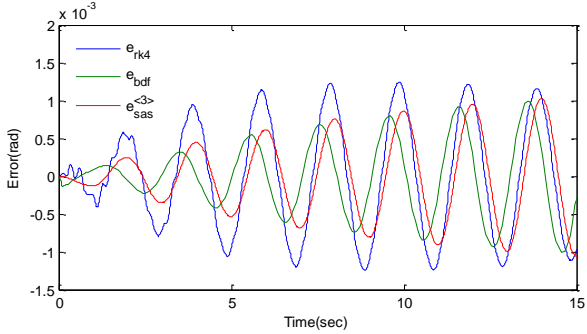

Fig. 4. Error curves of RK4, BDF and 3rd order SAS

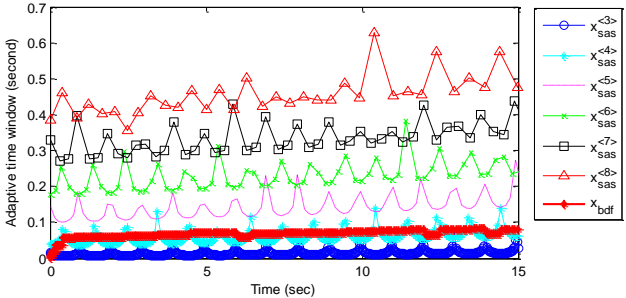

Fig. 5. Adaptive time windows from BDF and SASs with orders from 3 to 8

*B. Performance Tests on SAS Derivation*

This subsection studies the offline performance in deriving the proposed time-power series based SAS in terms of the time consumption and required space for storage. To achieve a fair comparison with the ADM [13], in this subsection, (i) a power system model only containing 6th order generators is used; (ii) the dynamic bus approach, which further reduces the complexity of the SAS, is not used in the proposed method; and (iii) all variables and parameters are symbolized when deriving the SAS, including time $t$, initial state variables, reduced admittance matrix.

The time cost for the ADM to offline derive the 2nd order SAS of the New England 39-bus 10-machine power system is about 16339.7s and this solution stored in text without compression takes about 15380KB space. On an Inter Core™ i7-6700 3.4GHz desktop computer using Symbolic Math Toolbox in Matlab, deriving the 2nd order time-power series based SAS only costs about 4.4s. The solution recorded in text is much more compact and only takes about 307 KB space, i.e. 2.0% of the space by ADM. In addition, the time costs for deriving higher-order time-power series SASs are shown in Table III.

TABLE III
TIME COSTS FOR DERIVING HIGHER-ORDER SASs

| Order of SAS | 3 | 4 | 5 | 6 | 7 | 8 |
|---|---|---|---|---|---|---|
| CPU time (sec) | 10.1 | 22.5 | 52.6 | 116.3 | 246.6 | 486.0 |

*C. Performance Tests on SAS Evaluation for Polish 2383-bus system*

The Polish power system model adopted here has 327 machines, 2383 buses and 1826 loads [24]. All generators use the model in (13). 20% of load buses, i.e. 366 buses, have motor loads that consume 40% of the total active power load on those buses. The non-motor loads on those 366 buses and the rest 1460 buses are represented by a ZIP load model with 20% constant impedance, 30% constant current and 50% constant power components. The contingency under our tests is a three-phase temporary fault on generator bus 10, the same location as [17]. The fault is cleared after 4 cycles without disconnecting any line, and the post-fault system is simulated for 10 seconds to cover the period of transient dynamics. To have a reference for comparison, the forward Euler (FE) numerical integration method with fixed time windows is adopted as a fast simulation method, and two cases are created below. Note that any other numerical integration method can also be used here for the comparison. The SAS with $n$=2 and $n_v$=1 is used for the simulation with adaptive time windows. The error-rate tolerances for all rotor angles, all voltages and all mechanical powers are respectively selected as 2 degrees/s, 0.01 pu/s and 0.001 pu/s.

- Case 1: a small time window, i.e. 0.2 ms, is used in the FE method to produce a reference result for checking the accuracy of the SAS based simulation.
- Case 2: a 1ms time window is used to provide a typical time cost to show the improvement achieved by the SAS based approach.

Because each case generates simulation results on 327 generators and the results on many buses. Here, select five generators and five motors near the fault location to show their simulation results in Fig. 6a-6h, 6i-6k and 6l, which depict all eight states of the generators, all three states of the motors and their terminal bus voltages. From Fig. 6, the results from the SAS based simulation match well the reference results from the FE method. Among all 327 generators, the largest angle difference between the results of the SAS based approach and the FE method is less than 0.7 degree, which is much smaller than the maximum error determined by the error-rate tolerance, i.e. 20 degrees. It verifies the conservativeness of the proposed adaptive time window. These figures show that the SAS based approach is able to simulate a power system in DAEs with adaptive time windows satisfying the given error-rate tolerances.

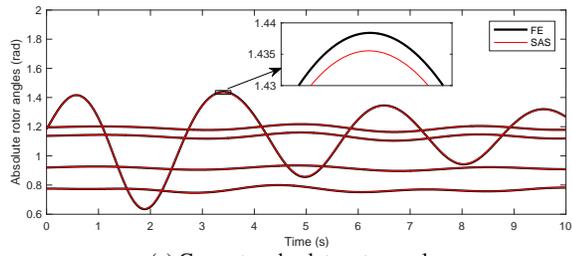

(a) Generator absolute rotor angle

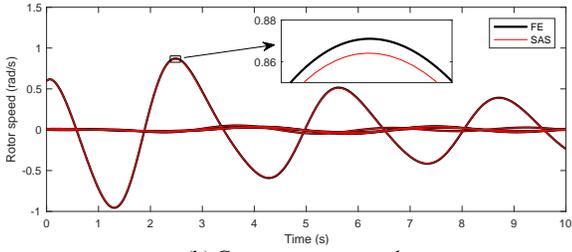

(b) Generator rotor speed

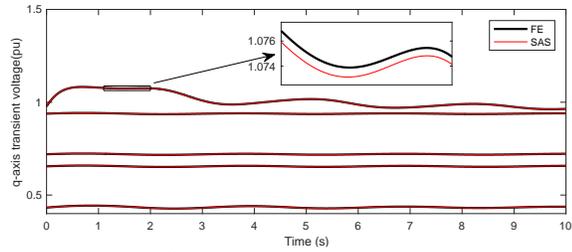

(c) Generator q-axis transient voltage

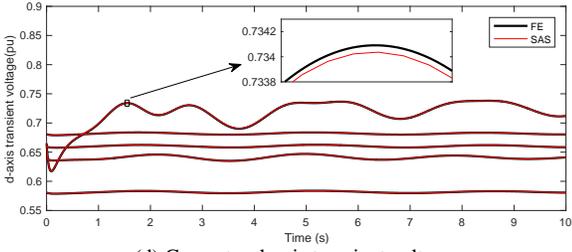

(d) Generator d-axis transient voltage

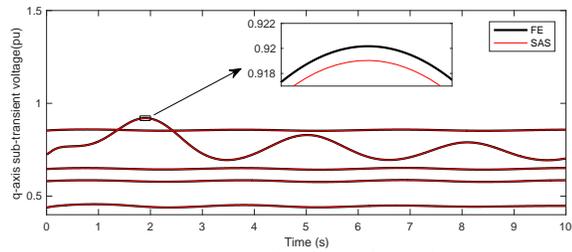

(e) Generator q-axis sub-transient voltage

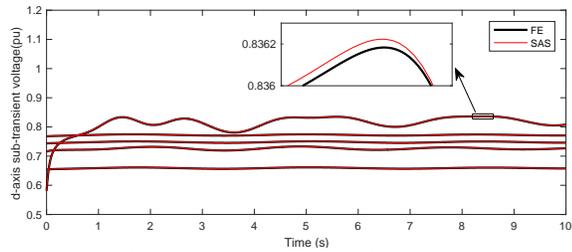

(f) Generator d-axis sub-transient voltage

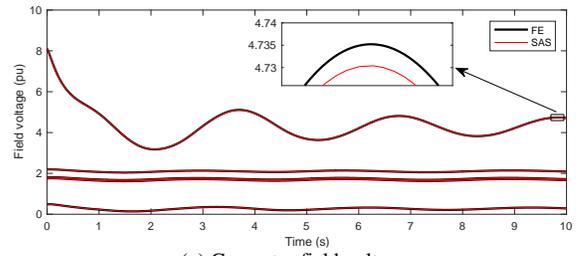

(g) Generator field voltage

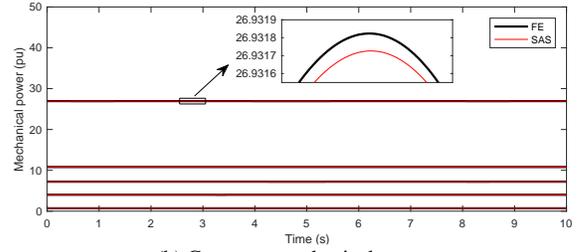

(h) Generator mechanical power

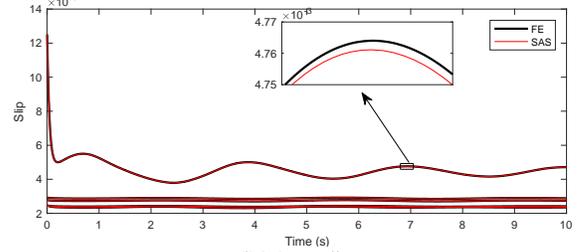

(i) Motor slip

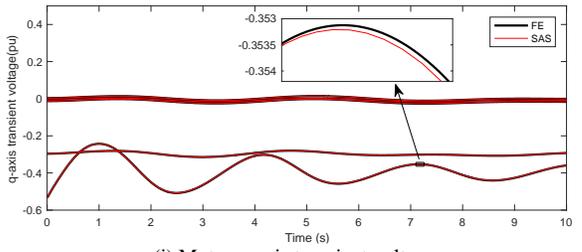

(j) Motor q-axis transient voltage

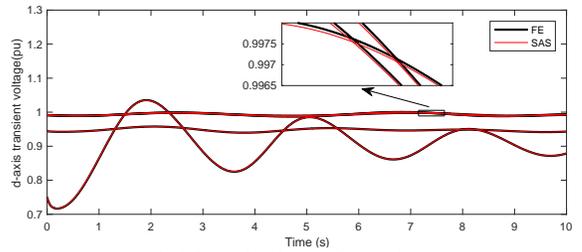

(k) Motor d-axis transient voltage

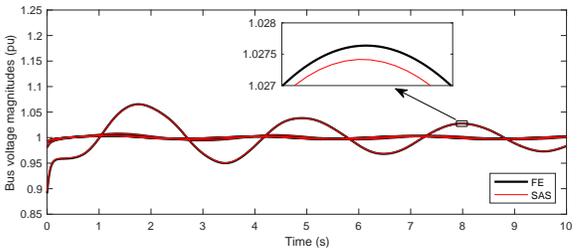

(l) Bus voltage magnitudes

Fig. 6. Simulation results of SAS based approach and the FE method

Table IV summarizes the comparison between the SAS based approach and the FE method when simulating case 2.

For the FE method, most of the time is spent on solving the network AEs, which has to be performed in each time window. Since the fixed time window is small, i.e. only 1ms, then the AEs have to be solved for many times, leading to a total of 54.6s time cost for simulating the 10-second dynamics of the system. In the SAS based simulation without considering parallel computing, denoted as "SAS-Sequential", the time spent on solving network AEs is significantly reduced to 6.11s, since the AEs are solved for much fewer times with the longer adaptive time windows. As shown in Fig. 7, the adaptive time windows determined by the SAS based simulation can be as large as 10ms-18ms and the average is about 12.8ms. If all SAS evaluations are performed by a single processor in a sequential manner, that takes as much as 153.9s. Thanks to the fact that the SAS evaluations can be parallelized among multiple processors, if we can afford one processor for each dynamic element, i.e. 693 processors in total for all generators and all motors, and envisage an ideal parallelization, the total time cost of SAS evaluations (i.e. simulation) will decrease from 153.9s to 0.44s and the overall time cost of the SAS based simulation will be less than 7s, which is much less than the 54.6s by the FE method. In addition, the SAS evaluation of each dynamic element can be further parallelized if using an ideally tremendous number of processors, i.e. 12.6 million, the overall time cost can be further reduced to about 6.5s in theory. Since that incremental improvement is too expensive, the parallelization is recommended for the element level to achieve a high payback.

TABLE IV
COMPARISON BETWEEN SAS BASED APPROACH AND FORWARD EULER

| Simulation routines | FE | SAS – Sequential | SAS – Ideal parallelization |
|---|---|---|---|
| Generator DEs | 5.1s | 65.9s | 0.20s |
| Motor DEs | | 88.0s | 0.24s |
| Network AE | 49.5s | 6.11s | 6.11s |
| Adaptive time window | N/A | 0.36s | 0.02s |
| Total time cost | 54.6s | 166.9s | 6.93s |
| # of time windows | 10000 | 781 | 781 |

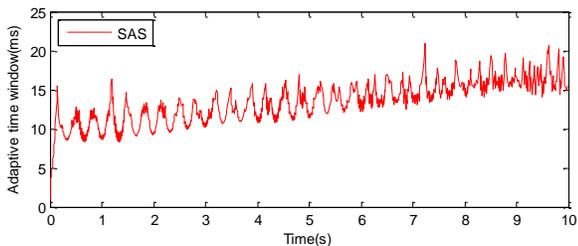

Fig. 7. Adaptive time windows determined in the SAS based simulation

Considering that the FE method in the above comparison solves network AEs fewer times, too, on a longer fixed time window, we test the FE method additionally with longer time windows. When using a 2ms fixed time window, the simulated rotor speed by the FE method is shown in Fig. 8, where one generator diverges. Because the simulation using 1ms time step gives all stable results, such a divergence is caused by numerical instability. Thus, the 2ms time window is too large for the FE method.

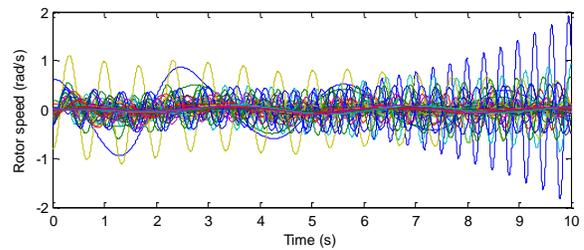

Fig. 8. Simulated rotor speed by FE with 2ms fixed time window

## V. CONCLUSIONS AND FUTURE WORK

This paper proposes a time-power series based semi-analytical approach for power system simulation as an alternative approach of the traditional numerical integration based simulation. The offline derivation of an SAS is faster than existing methods using the Adomian Decomposition. A dynamic bus method is proposed to extend the SAS based simulation from DEs to DAEs. An error-rate upper bound is proposed and used to determine an adaptive time window for the SAS. By applying adaptive time windows, the SAS based simulation eventually takes fewer time windows to finish any specific simulation period such that the most time-consuming task, i.e. solving the network AEs, is solved much fewer times. Leveraged by parallel computing, the time cost for SAS evaluations can be largely reduced such that the overall time cost for the SAS based simulation is significantly reduced compared to the forward Euler numerical integration method. Case studies performed on New England 39-bus system and Polish 2383-bus system show that the SAS based time domain simulation has potentials in online simulations.

## VI. ACKNOWLEDGEMENTS

The authors would like to thank Mr. Hantao Cui from the University of Tennessee, Knoxville for his suggestions on preparing the Polish system model in PSS\E format.